\documentclass[aps,pra,twocolumn,superscriptaddress,10pt]{revtex4-2}

\usepackage{amsmath, amssymb}
\usepackage{graphicx}
\usepackage{enumitem}
\usepackage[hidelinks]{hyperref}

\begin{document}
\title{Quantum Time-Series Learning with Evolutionary Algorithms}
\author{Vignesh Anantharamakrishnan}
\affiliation{Sorbonne Universit\'e, Pierre et Marie Curie Campus, 4 Pl. Jussieu 75005 Paris, France}
\author{Márcio M. Taddei}
\affiliation{ICFO-Institut de Ci\`encies Fot\`oniques, The Barcelona Institute of Science and Technology, 08860 Castelldefels (Barcelona), Spain}

\begin{abstract}
Variational quantum circuits have arisen as an important method in quantum computing. A crucial step of it is parameter optimization, which is typically tackled through gradient-descent techniques. 
We advantageously explore instead the use of evolutionary algorithms for such optimization, specifically for time-series forecasting. 
We perform a comparison, for diverse instances of real-world data, between gradient-descent parameter optimization and covariant-matrix adaptation evolutionary strategy. 
We observe that gradient descent becomes permanently trapped in local minima that have been avoided by evolutionary algorithms in all tested datasets, reaching up to a six-fold decrease in prediction error. Finally, the combined use of evolutionary and gradient-based techniques is explored, aiming at retaining advantages of both. 
The results are particularly applicable in scenarios sensitive to gains in accuracy.

\end{abstract}

\maketitle

\section{Introduction}
Machine learning has risen to tremendous prominence in the last decade with advances in many different tasks like classification, transformation, data analysis and generation in fields as diverse as customer segmentation, computer vision, sound and text interpretation, among many more.
At the same time, the breakthrough potential of quantum computing, together with recent advances in available quantum hardware, make it a suitable alternative for probing new techniques in computing tasks, including those of machine learning. 
In the era of noisy intermediate-scale quantum (NISQ) devices, variational quantum algorithms (VQAs)~\cite{Cerezo2021,Bharti2022} are an important candidate for such advancements for their current feasibility, as well as for being a testbed for techniques for future implementations.

VQA requires, as does classical machine learning, the optimization of many parameters, and the most common optimizations techniques used in VQA are gradient descent in its many refinements~\cite{Romero2018,Bharti2022}. However, evolutionary strategies have also been proposed and applied to certain tasks, such as finding molecular or many-body ground states~\cite{Rattew2020,Chivilikhin2020,Robert2021,Wakaura2021,Wakaura2021a,Anand2021,Huang2022,Chen2022d,Franken2022,Schleich2023,Friedrich2023}, quantum encoding and state preparation~\cite{Chen2022,Creevey2023,West2023}, as well as classically defined tasks~\cite{Zhao2021,Pellow-Jarman2023,Turati2023,Albino2023,Kolle2024}.
The exploration of different optimization techniques is necessary, among other reasons, because of barren plateaus, which can easily affect VQA optimization~\cite{McClean2018,Wang2021,Cerezo2021b,Ragone2023}. Evolutionary algorithms do not \emph{per se} eliminate the issue of barren plateaus~\cite{Arrasmith2021}, but as a different  approach to optimization, offer different possibilities to navigate them. Another common problem in VQA parameter optimization is the convergence to local minima; in this regard evolutionary algorithms are particularly welcome because of their global nature. We shall see that the techniques proposed and their efficacy in our tests precisely suggest that avoiding local minima is a crucial advantage of evolutionary algorithms in VQA over gradient descent.

One kind of problem where VQAs can be applied to is that of time series. Time series are data that depend on a one-dimensional parameter with a sequential ordering (time or a variable akin to time). They can represent myriad kinds of information, such as economics indicators, oral and written language, weather data, etc. The supervised learning task we will tackle is forecasting, the ability to predict later points in a series based on the previous ones. It has an expected structure --- closer points being typically more relevant to prediction than distant ones ---, something reflected in the architecture for machine learning for forecasting tasks. There is widespread interest in advancing the capabilities of time-series forecasting, be it in accuracy or in predicting further into the future.

Here we study the use of evolutionary algorithms against gradient descent to optimize parameters in a VQA applied to time-series forecast, with a focus on scenarios where gradient-descent methods underperform in terms of accuracy. Such scenarios can be naturally brought about by extending how far into the future predictions are made, as it readily and realistically increases errors. 
The evolutionary techniques presented are able to achieve up to a six-fold decrease in error given the same quantum-circuit ansatz and dataset. 
Since evolutionary algorithms have long training times, 
we also present a hybrid method, combining gradient descent and evolutionary training to obtain an even higher accuracy while mitigating the longer train time. In fact, the hybrid method reaches an almost ten-fold decrease in error compared to gradient descent in one instance. 
In Section \ref{sec:prelim} we present the methods used in our experimentation, whose results and discussion are found in Section \ref{sec:results}. Section \ref{sec:conclusions} is left for concluding remarks.

\section{Methods}
\label{sec:prelim}

\subsection{Variational Quantum Algorithms}

VQAs are hybrid algorithms which employ a quantum circuit with variable, parameter-dependent gates and a measurement in the end that produces a classical variable, from which a prediction is derived. As in classical machine learning, training is used to classically optimize the parameters such that the predictions match the ground truth of the training data (minimize a suitable cost function). Two important differences of VQA compared to classical methods are that the measurement outcomes are intrinsically probabilistic (differently from classical methods that introduce randomness, such as mini-batching) and that evaluating the derivatives of the cost function requires additional runs of the circuit (typically using the parameter-shift rule~\cite{Mitarai2018, Schuld2019a}).

The simplest way to perform such optimization is by gradient descent, which follows the direction of steepest descent of the cost function. 
It has the advantage of a speedy calculation and is considered the primary method for parameter optimization in classical and quantum algorithms. One of its challenges in VQA is the issue of \textit{barren plateaus}~\cite{McClean2018}, where derivatives of a cost function  are exponentially suppressed in an exponentially large fraction of the parameter space. The issue affects any optimization method in VQA~\cite{Arrasmith2021}, but global methods probe the parameter space differently, hence may provide different behaviors in the face of barren plateaus.

More importantly for the present work, gradient-descent methods are local, hence subject to convergence to local minima. In spite of the existing techniques to avoid them, we shall see that, indeed, the behavior of the gradient descent is consistent with being trapped in local minima that can be otherwise circumvented. 

In the experimentation of this paper, gradient descent will be performed with the Adam optimizer~\cite{Kingma2017} with with a learning rate of 0.03, specifically the PyTorch implementation~\cite{Paszke2019}. All other parameters in the Adam optimizer were set to their \texttt{torch} default values.

Evolutionary methods for optimization have a rich history and avail of a wide range of different algorithms~\cite{Back1993,Back2023}. For our case of optimizing multiple real parameters in a quantum circuit, we have chosen the Covariance Matrix Adaptation Evolution Strategy (CMA-ES)~\cite{Hansen2003,Hansen2023}. It samples from a multivariate normal distribution specified by a mean and a covariance matrix, which are taken from their previous-generation values but, crucially, with weight factors to enhance the contribution of better solutions. 
As for the CMA-ES implementation, we shall use the Python \texttt{cma} package version 3.3.0
with the  \texttt{cma.CMAEvolutionStrategy} class, with a population size of 10, a randomized initial (multivariate) mean and an initial standard deviation of 0.5 for all weights.

Because our main focus is the comparison between different strategies of the classical optimization step of VQAs, the quantum circuits involved shall be classically simulated. This will be done on the \texttt{qiskit} 1.1.0 package coupled with the Qiskit IBM-Runtime version 0.23.0 which uses the parameter-shift rule by default. The laptop we used for benchmarking has an AMD Ryzen 9 6900HS with an RTX 3070Ti and 16GB of RAM. 

One may notice that the training is performed with few to no changes to the default hyperparameter values. This is in line with our goal of comparing the different types of optimization. The tuning of such values would have to be performed to a similar extent on both methods in order not to bias the comparison towards one or another; however, the hyperparameters play such different roles in gradient-based and evolutionary strategies that there is no clear way to compare the tuning efforts of the two. As such, we leave almost all hyperparameters to their defaults. One could make the indirect argument that the \texttt{PyTorch} package is developed by such a larger list of contributors than the \texttt{cma} package that the defaults of the gradient-based optimization are arguably the fruit of much more extensive tuning than those of the evolutionary strategy. This would amount to a bias favoring the former, which does not run counter to our objective of showing advantages in the use of the latter.

\subsection{Time-series ansatz}
\label{sec:timeseries}
Time series comprises data that depend on a sequential one-dimensional variable, typically (though not necessarily) interpreted as time. 
For each timestep $t\in\mathbb N$, the input data is $x^{<t>}$, which is the value of the $t$-th datapoint. A common goal is forecasting, i.e., predicting later points based on previous ones. For that we aim to have the output $y^{<t>}$, which is the output of the learner using the input up to $x^{<t>}$, to reproduce the following datapoint $x^{<t+1>}$, i.e.\ $y^{<t>}\approx x^{<t+1>}$. For the case of forecasting several timesteps into the future, an iterative calculation is made. For instance, if data up to $t_0$ is to be used to predict datapoints up to $t_0+\Delta t$, then $y^{<t_0>}$ is used as input in lieu of $x^{<t_0+1>}$ to obtain $y^{<t_0+1>}$, which should be a good approximation to $x^{<t_0+2>}$. This is repeated iteratively 
until the prediction for the desired timestep $t_0+\Delta t$ is reached. This naturally accumulates errors when going further in time, and for that reason time-series problems rapidly become exceedingly difficult by increasing the amount of timesteps of the forecast into the future. Importantly, such increase in timesteps is by no means an artificial requirement, since predicting further into the future draws interest in most applications.

Classically, recurrent neural networks (RNN) are mostly used to tackle these problems. Its main component is the recurring unit, which takes in the current input datapoint $x^{<t>}$, together with a memory variable that encodes information about previous datapoints, and from that generates a prediction $y^{<t>}$ (as well as the next value of the memory variable). There are many different ways to classically compose the information from $x^{<t>}$ and the memory to produce $y^{<t>}$, the best-known being the long short-time memory (LSTM)~\cite{Hochreiter1997} and the gated recurrent unit (GRU)~\cite{Cho2014a, Cho2014b}.

\begin{figure*}
    \centering
    \includegraphics[width=\textwidth]{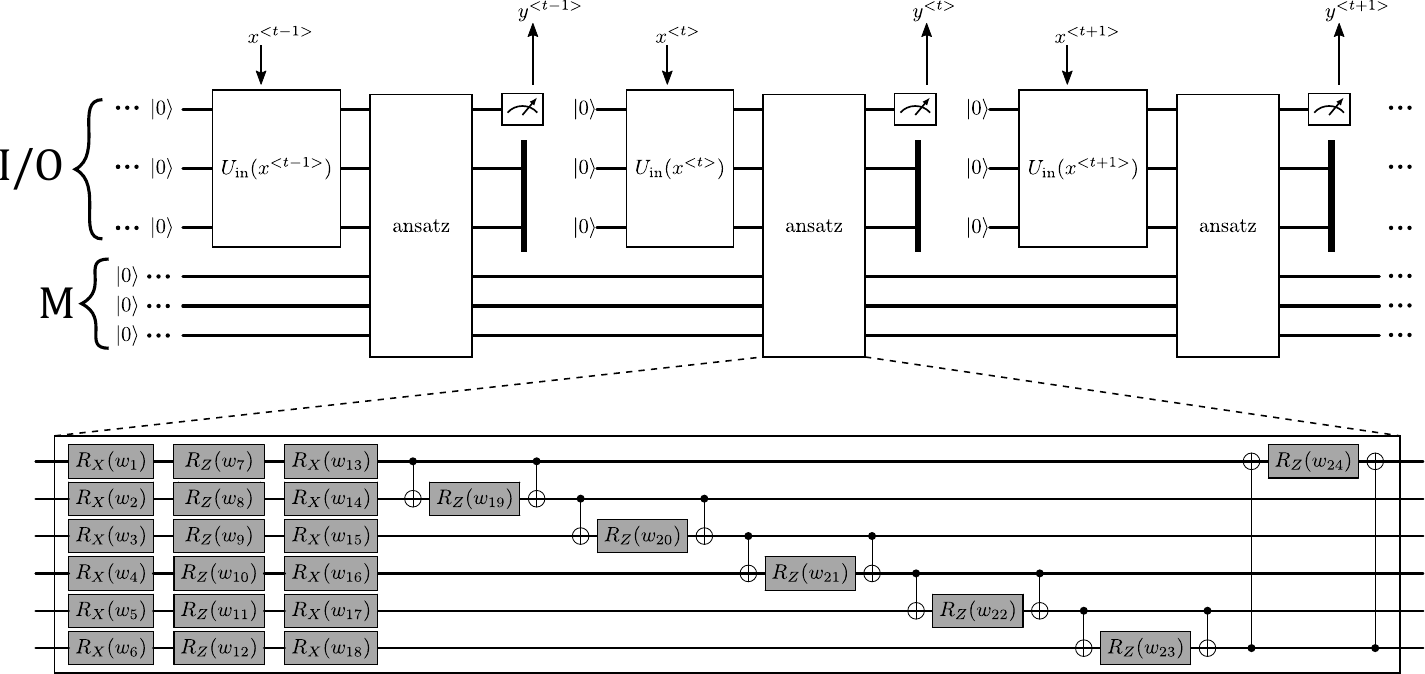}
    \caption{Quantum recurrent neural network (QRNN) circuit used in this work, after~\cite{Li2023}, and its recurring ansatz.}
    \label{fig:QRNNcircuit}
\end{figure*}

For our exploration of quantum neural networks, we will use, after~\cite{Li2023}, a quantum recurrent neural network (QRNN) that closely mimics the classical one, with qubits $I/O$ dedicated to encode the current input $x^{<t>}$ and later extract the output $y^{<t>}$, and qubits $M$ dedicated to act as memory of past inputs, see Fig.~\ref{fig:QRNNcircuit}. 
Angle encoding will be used to input the data, encoding $x^{<t>}$ three times, one in each $I/O$ qubit through an $R_Y$ gate; a single-qubit computational-basis measurement will generate the outputs $y^{<t>}$. 
Instead of LSTM or GRU recurrent blocks, we will have a parametrized quantum circuit, the choice of which is typically called a quantum-circuit ansatz. The ansatz used in this recurrent block, composed of three $I/O$ qubits and three $M$ qubits and 24 variational parameters in total, is depicted in Fig.~\ref{fig:QRNNcircuit}. Besides the initial one-qubit gates, the two-qubit gates cyclically link the qubits, allowing for information exchange between $I/O$ and $M$ qubits.
The code used to perform the simulations is available on \url{https://github.com/Viggi999999/EvoAlgoCode}.

\section{Results and discussion}
\label{sec:results}

For each dataset we will perform the experimentation in three stages, meant to probe gradient descent, CMA-ES and their combination. 
First, to establish the performance of gradient descent, in particular its accuracy limitations, we perform gradient descent for 100 epochs. This is an intentionally excessive number, and displays the plateau for the accuracy achieved through gradient descent only.
Second, we run CMA-ES for 11 epochs, an amount that will suffice to show the increased accuracy of CMA-ES. 
Third, we run the hybrid method, which consists in using the gradient-descent algorithm until it reaches a plateau, then switching for CMA-ES. We perform the switch after a fixed number of 20 gradient-descent epochs, which in our experimentation will be enough to reach this plateau.  Because we expect the hybrid method to reach a lower error with less training, we subsequently run only 9 CMA-ES epochs.

We apply our procedure to four univariate time-series datasets:
\begin{enumerate}[label=\roman*)]
    \item the daily gold-price series from the World Gold Council~\cite{Gold};
    \item the Santa Fe time series A, generated from an NH${}_3$ laser for the Santa Fe forecasting competition~\cite{Weigend1993};
    \item the Mackey-Glass dataset~\cite{MGdataset}, generated from a solution to the Mackey-Glass model~\cite{Mackey1977},
			\begin{equation}
				\frac{dx}{dt} = \beta x(t) + \frac{\alpha \ x(t-\tau)}{1+x^{10}\ (t-\tau)}
				\label{eq:MackeyGlass}
			\end{equation}
    with $\alpha = 0.2$, $\beta = -0.1$, $x(0) = 1.2$, and $\tau = 17$~\cite{Waheeb2018}; and 
    \item mean air-pressure data from daily weather data in Delhi, India~\cite{DelhiDataset}.\end{enumerate}
These constitute a diverse range of time-series data, and possess a chaotic behavior. 

For the purposes of our experiment, we use the first 100 datapoints of each dataset, split 80-20 into train and test sets, respectively. This is not a large amount of datapoints, which suits our goals  firstly because it is sufficient to draw meaningful conclusions and secondly because it leads to a somewhat low accuracy, allowing for more room for it to be improved by the evolutionary methods.  

\begin{figure*}
    \centering
    \includegraphics[width=\columnwidth]{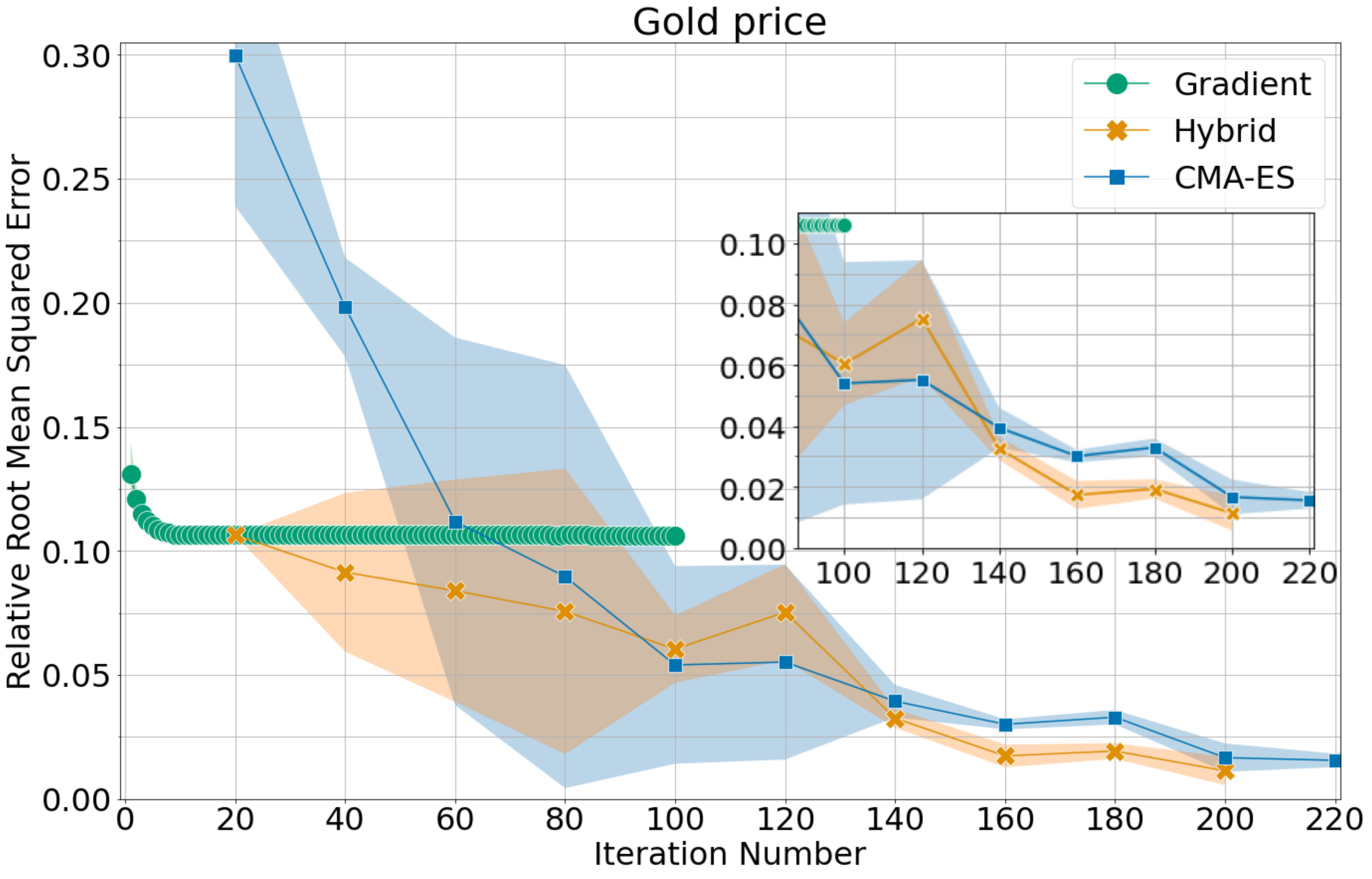}
    \includegraphics[width=\columnwidth]{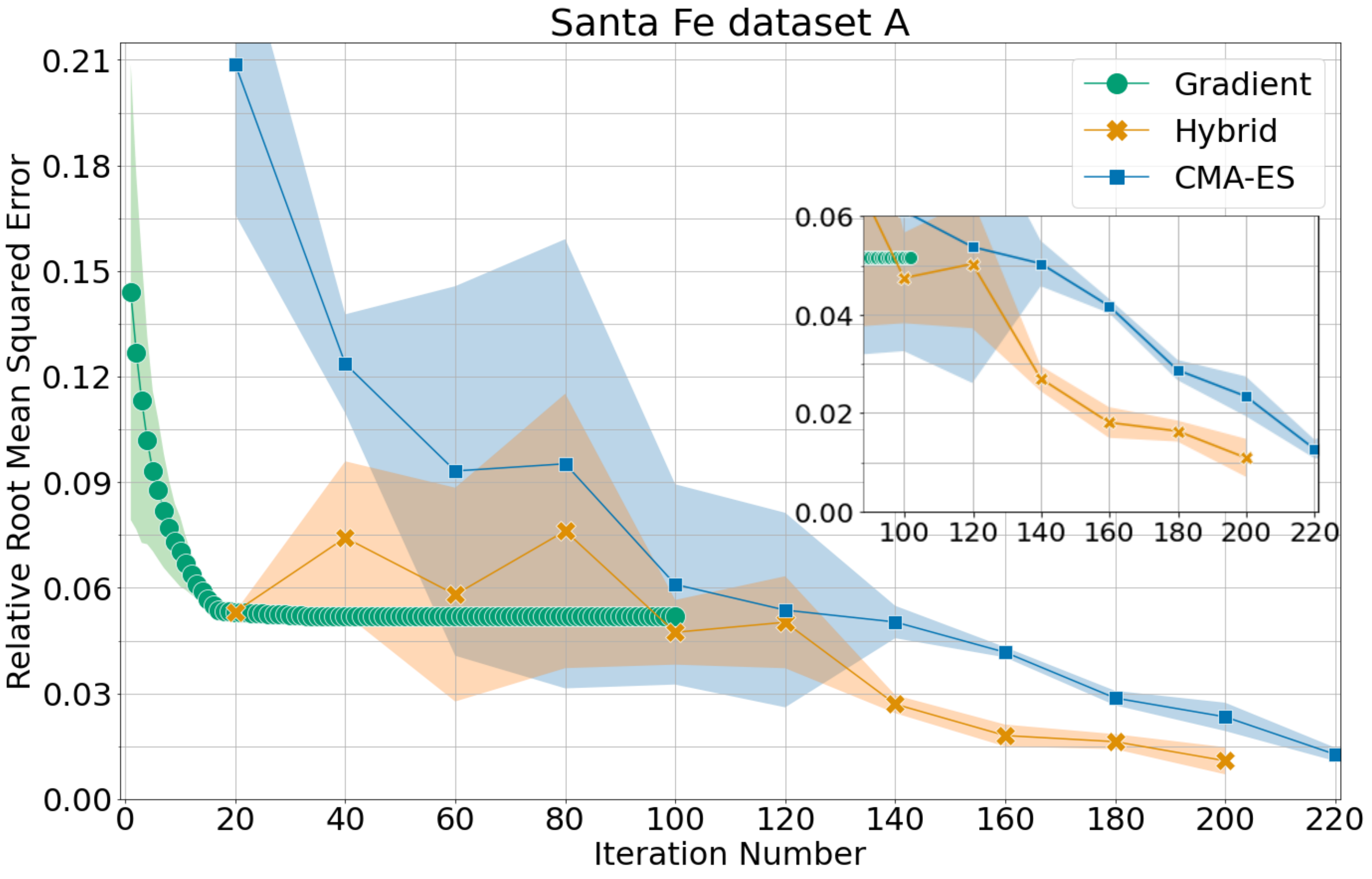}
    \includegraphics[width=\columnwidth]{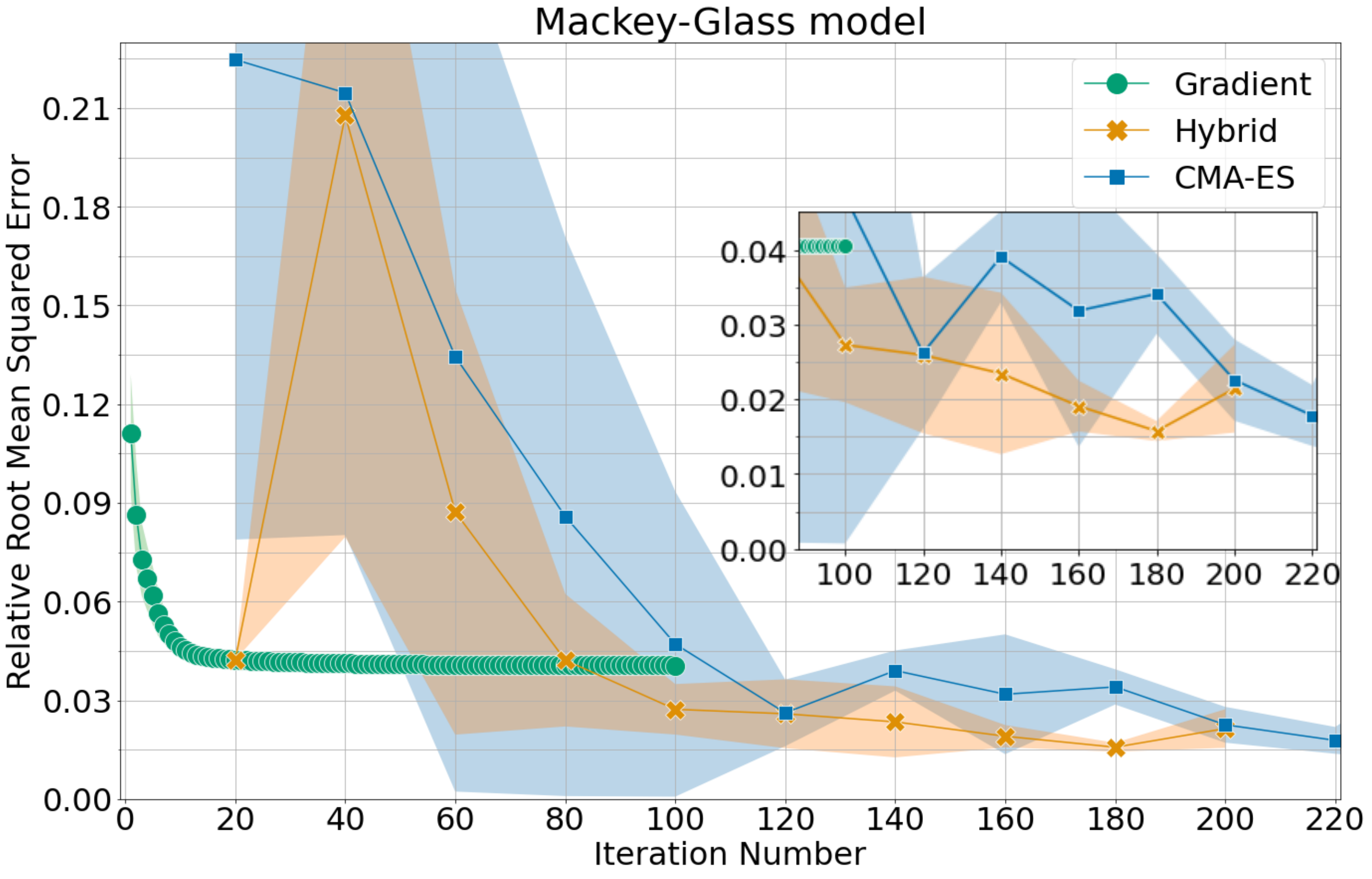}
    \includegraphics[width=\columnwidth]{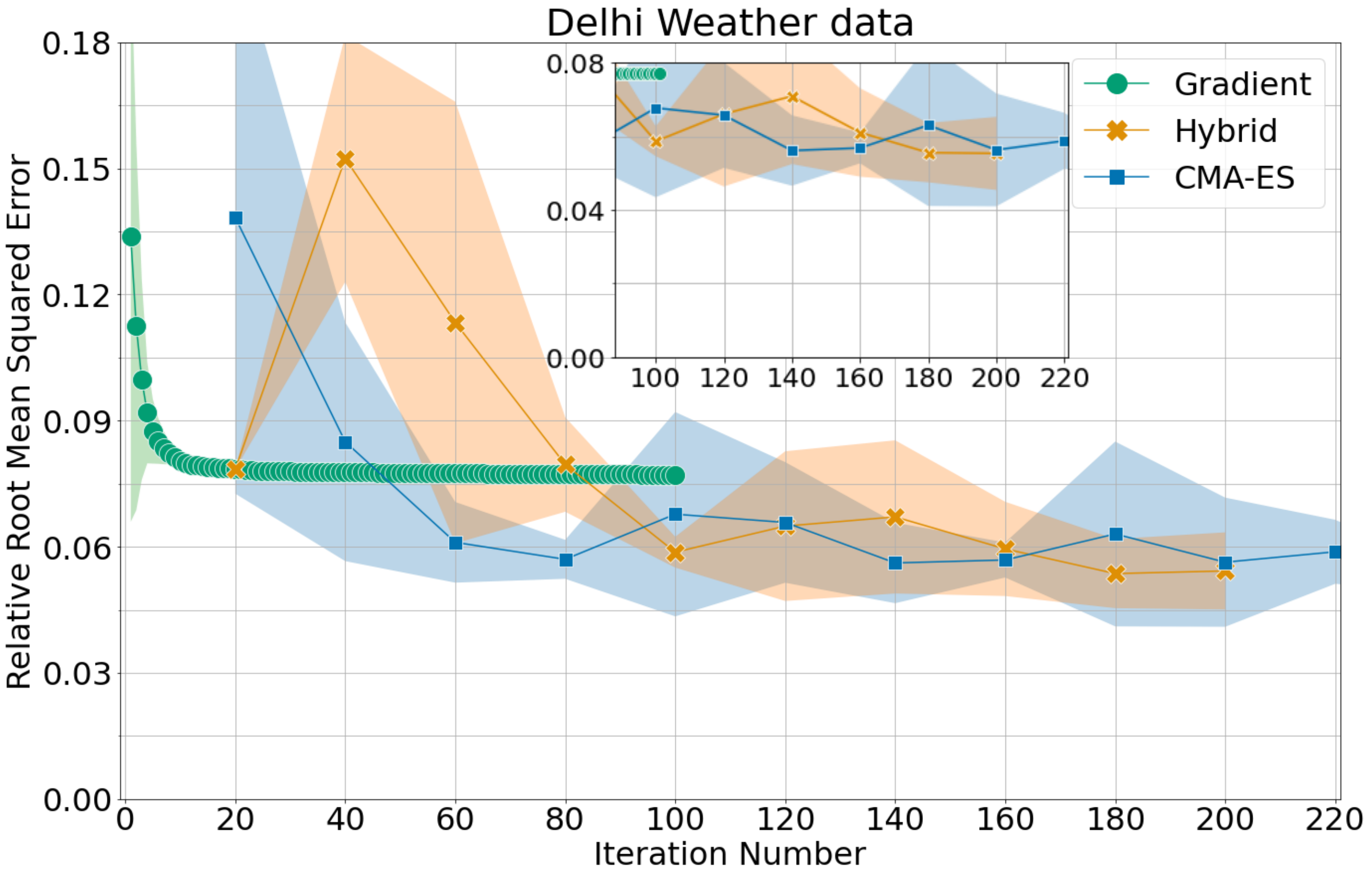}
    \caption{Forecast error for the gold-price, Santa Fe, Mackey-Glass, and Delhi-weather datasets, with predictions for 7, 1, 4, 9 timesteps into the future, respectively. For each dataset, three different optimization strategies are used: gradient descent (Adam), the evolutionary strategy CMA-ES, and their hybrid, which uses gradient descent as a warm start (the first point of each hybrid curve by definition coincides with the gradient-based value). The relative RMS error is shown, with points and shaded regions corresponding respectively to mean and one standard deviation after 5 runs. The x-axis values correspond to gradient-descent epochs; for CMA-ES and hybrid the points are plotted with a horizontal delay of 20 units to visually depict that they take circa 20 times longer to train, per epoch, than gradient descent.}
    \label{fig:mainresults}
\end{figure*}

The main results are in Fig.~\ref{fig:mainresults}, which plots the relative root mean square (RMS) error at each epoch, as mean and standard deviation after 5 runs. In all datasets, gradient descent very clearly reaches a minimum error plateau which it cannot surpass. We can safely interpret it as a local minimum at which the gradient-based method is stuck. 
We have adjusted the problem difficulty (through the number of prediction timesteps) for each dataset to have different levels of gradient-based accuracy to improve upon. Naturally, for each dataset the three methods (gradient, CMA-ES, hybrid) tackle the same task, with the same number of timesteps. 

\begin{table}[tb]
    \centering
     \begin{tabular}{|c|c|c|c|}\hline
      Dataset       & Gradient  & CMA-ES & Hybrid\\\hline
      Gold price    & 10.6\%    & 1.6\%  & 1.1\%\\
      Santa Fe      & 5.2\%     & 1.3\%  & 1.1\%\\
      Mackey-Glass  & 4.1\%     & 1.8\%  & 1.6\%\\
      Delhi weather & 7.7\%     & 5.6\%  & 5.4\%\\\hline
     \end{tabular}
    \caption{Lowest average relative RMS error on the forecast of each dataset, per optimization method. Equivalently, lowest point in each plot in Fig.~\ref{fig:mainresults}.}
    \label{tab:lowest}
\end{table}

Employing CMA-ES was effective in breaking through the plateau on gradient descent, with varying degrees of improvement. 
As shown in Table~\ref{tab:lowest} and Fig.~\ref{fig:mainresults}, a six- and four-fold reduction in error was found for the gold-price and Santa Fe datasets, respectively, when the gradient optimization was replaced by the CMA-ES method. For the Mackey-Glass dataset the error was halved, and a more modest reduction of one quarter was found for the Delhi weather. This substantiates the claim that the evolutionary strategy is able to escape from local minima at which gradient descent (Adam) was stuck. None of the strategies can be assumed to have reached a global minimum, but the CMA-ES values are considerably lower than the gradient-based ones.

The use of the hybrid method presented a moderate increment in accuracy over CMA-ES only, as seen in Table~\ref{tab:lowest}, which allowed for an almost ten-fold error reduction when compared with gradient optimization for the gold-price dataset. The more consistent advantage obtained from using the hybrid method, however, was that these low error values were obtained with fewer epochs than with CMA-ES only.
For the gold-price, Santa Fe, and Mackey-Glass datasets, one can see that certain error values were attained 3 epochs earlier (out of 11) with the hybrid method than with CMA-ES. As such, the hybrid method could often mitigate the intrinsic disadvantage of longer evolutionary-algorithm training times for all but one dataset.

Comparing the different datasets, we see that an important factor  for a sizable error reduction through evolutionary strategies was a larger gradient-based error, i.e.\ more room for improvement. The exception to this was the Delhi weather dataset. For this data composed of daily air-pressure values, Fig.~\ref{fig:mainresults} shows that the evolutionary strategies were eventually stuck in local minima with limited improvement from the gradient-based method. Optimization of CMA-ES hyperparameters would be the natural candidate to overcome this limitation but, to preserve a fair comparison between gradient-based and gradient-free methods, we have opted not to perform it.

\section{Conclusions and Perspectives}
\label{sec:conclusions}

In this work, we have probed evolutionary methods to train parameters in parameterized quantum circuits. This was done in the task of forecasting future data in time-series problems, tackled by quantum recurrent neural networks. The CMA-ES was chosen as evolutionary strategy, as it is particularly suitable for tasks with continuous numerical outputs. Through application on a diverse series of datasets, we have seen that evolutionary strategies have been able to escape local minima that gradient-based (Adam) methods were stuck at, often reaching errors many times smaller. We have also proposed a hybrid method, that uses gradient descent up to its plateau as a warm start to evolutionary optimization, obtaining an increment in accuracy relative to CMA-ES with less training.

We conclude that the strategies shown here are particularly useful when the accuracy of gradient-based methods is unsatisfactory. This can happen rather naturally in time-series forecast through a combination of factors such as increased number of prediction timesteps, data scarcity, high accuracy goals.

Let us mention further explorations of the topic. One possibility involves different regimes for hybrid training, exploring the advantages of gradient and evolutionary training. A second one is the combination of evolutionary training and evolutionary ansatz updating~\cite{Rattew2020,Huang2022,Franken2022,Albino2023,Pellow-Jarman2023,Schleich2023}. This would involve different evolutionary strategies, since ansatzes form a discrete set, and the interplay between discrete and continuous evolutionary strategies is of interest.

\vspace{.8cm}

\begin{acknowledgments}
This work was supported by the Government of Spain (Severo Ochoa CEX2019-000910-S, FUNQIP, Misiones CUCO Grant MIG-20211005, European Union NextGenerationEU PRTR-C17.I1) and European Union (PASQuanS2.1, 101113690), Fundació Cellex,
Fundació Mir-Puig, Generalitat de Catalunya (CERCA program).
\end{acknowledgments}

\bibliography{CUCO,extra}

\end{document}